# Concerning the conjugation of field-aligned currents.

O.V. Mager, P.A. Sedykh, E.A. Ponomarev, and A.V. Tashchilin
Institute of Solar-Terrestrial Physics SB RAS, Irkutsk, Russia
pvlsd@iszf.irk.ru

**Abstract.** It is known that the combined action of convection and pitch-angle diffusion is responsible for the formation of gas pressure distribution in the magnetosphere [3]. Plasma pressure, in turn, determines - within the framework of a given magnetic field model - the density of bulk currents in the magnetosphere. With a knowledge of the bulk currents as a function of coordinates, we can calculate the field-aligned currents as a divergence of bulk currents. On the other hand, specifying the convection model is equivalent to specifying the electric field model. Since within the approximation of equipotential field lines the electric field is common to the magnetosphere and ionosphere, bulk currents and field-aligned currents in the ionosphere can be formally calculated subject to the condition that ionospheric conductivity is wholly determined by electron precipitation from the magnetosphere. The precipitation intensity is readily inferred from the same magnetospheric model. Thus we have two systems of field-aligned currents. One system is calculated from the model of plasma pressure distribution in the magnetosphere, and the other is inferred from a given model of the electric field and the electroconductivity model calculated from electron precipitation. This brings up the question: How can these two systems of field-aligned currents be reconciled? From previous studies [4,5] it is known that magnetospheric convection "adjusts itself" to the level of energy losses in the ionosphere. Based on this, an attempt can be made to achieve a conjugation of the aforementioned two systems of field-aligned currents. This paper is devoted to analyzing such an attempt.

**Introduction**

Over 30 years ago, C.F.Kennel suggested the idea [2] that the pitch-angle diffusion of particles into the loss cone, together with adiabatic compression of plasma during the convection into the magnetosphere determines the behavior of the contents of the magnetic flux tube, the plasma tube. Gas pressure builds up under the action of an adiabatic compression of plasma drifting in a magnetic field with increasing strength and precipitation-induced losses. A combined effect of these factors leads to an expression for gas pressure in the form [3]:

$$P = P_\infty(t)\left(\frac{L_\infty}{L}\right)^{\frac{20}{3}} \cdot \exp\left[-\int \frac{dt}{\tau}\right]; \quad \frac{dt}{\tau} = \frac{dR}{V_R \tau} = \frac{R d\lambda}{V_\lambda \tau},$$

$\tau$ - is the characteristic time of pitch-angle diffusion of protons and electrons; V – is the convection velocity; P – is gas pressure.

In this case the magnetosphere develops a plasma pressure distribution, such as shown in Fig. 1.

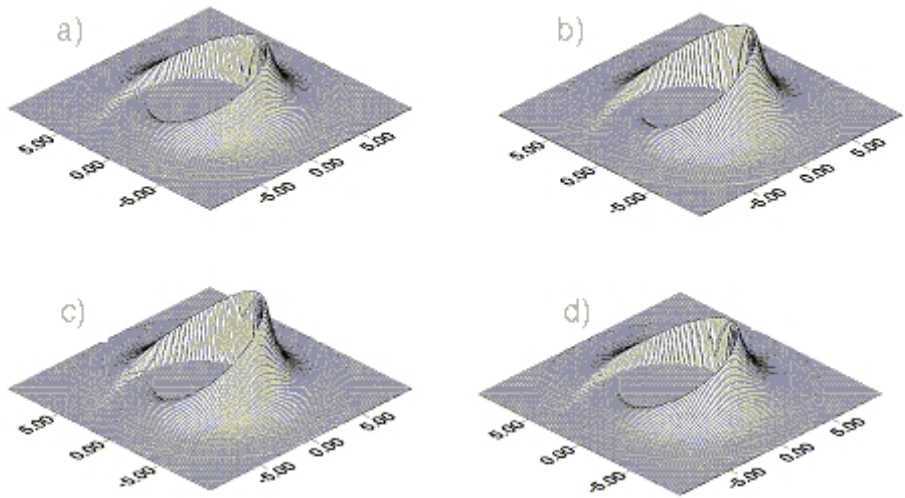

*Fig. 1. Gas pressure relief for steady-state boundary conditions but varying electric field of the convection $E_0$: a) t=0 s, b) t=1000 s, c) t=2800 s, d) t=4500 s. t - current time in the model from [3].*

In these calculations, boundary conditions are time-independent, that is, the supply of plasma through the boundary remains uniform, and only the convection velocity changes. It is seen how the pressure peak increases with the increasing electric field and how it reverts to the original level. Noteworthy is the relatively gentle backward slope of the relief and a very steep forward slope. If the boundary conditions are unsteady and pressure increases for a certain time on the boundary of the region, from which plasma starts, the resulting plasma "bunch" drifts downstream of the convection. Since the resulting pressure amplitude is the product of the undisturbed signal amplitude by the magnitude of the disturbance, there emerges the picture shown in Fig. 2. The sequence in Fig.2 illustrates the pressure relief time history as a function of the convection electric field and nonstationary boundary conditions varying as:

$$E_0 = E_{00} \begin{cases} 1; & t < -900s \\ 2t/1800 + 2; & -900s \leq t \leq 900s \\ 1; & t > 5400s \end{cases} ; \quad G = G_0^a \begin{cases} 1; & t' < 0s \\ 2t'/500 + 1; & 0s \leq t' \leq 500s \\ -t'/500 + 4; & 500s < t' \leq 1500s \\ 1; & t' > 1500s \end{cases}$$

$$P = G^a(t')A^a(R(t))$$

The function $G(t')$ thus represented a disturbance "drifting" downstream the convection with the drift velocity of this plasma component.

One can see that at the time when the plasma bunch reaches the region of maximum steady-state pressure, there occurs a powerful short-duration outlier of plasma pressure (density and, accordingly, intensity of particle precipitation). We interpret it as break-up substorms [3].

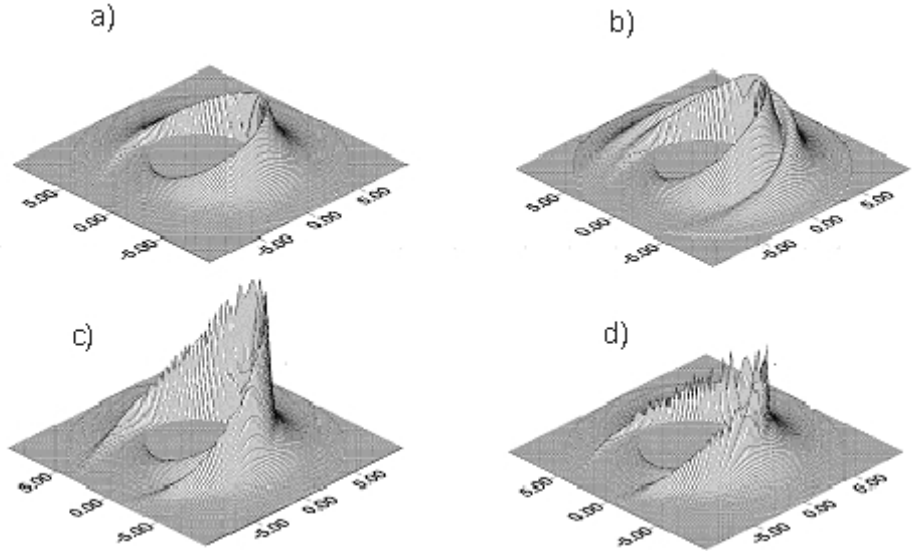

*Fig. 2. Gas pressure relief for unsteady boundary conditions:
a) t=0 s, b) t=1000 s, c) t=2800 s, d) t=4500 s.*

Since in plasma with isotropic pressure (such as observed in the magnetosphere within $3.5 < L < 7$, where L is the McIlwain parameter) the plasma pressure relief wholly determines the density of bulk currents for particles with the energy less than 15 keV, then:

$$\mathbf{j}_\perp = c\,[\mathbf{B} \times \nabla p_g]/B^2 , \qquad (1)$$

where **B** is the magnetic field strength, **p**$_g$ is gas pressure, and **c** is the velocity of light.

**Formulation of the problem and computational technique**

The flux density of precipitating particles is rather sharply localized in the space and produces at the Earth a clearly pronounced precipitation oval (Fig. 3).
Sharp spatially localized regions of increased conductivity in the ionosphere also correspond to such a distribution of precipitation. In the energy range 1-20 keV in which auroral electrons exist, about 30 electron-volt of the flux energy supplied to the ionosphere are expended in producing an electron-ion pair. Hence the ionization rate is ~ $j_\varepsilon/H\delta\varepsilon$ ($\delta\varepsilon$ in ergs), where $j_\varepsilon$ is the energy flux density of precipitating electrons (ergs/cm), and the electron density in steady-state conditions:

$$n_e \sim (j_\varepsilon/H\delta\varepsilon\alpha)^{1/2} \qquad (2)$$

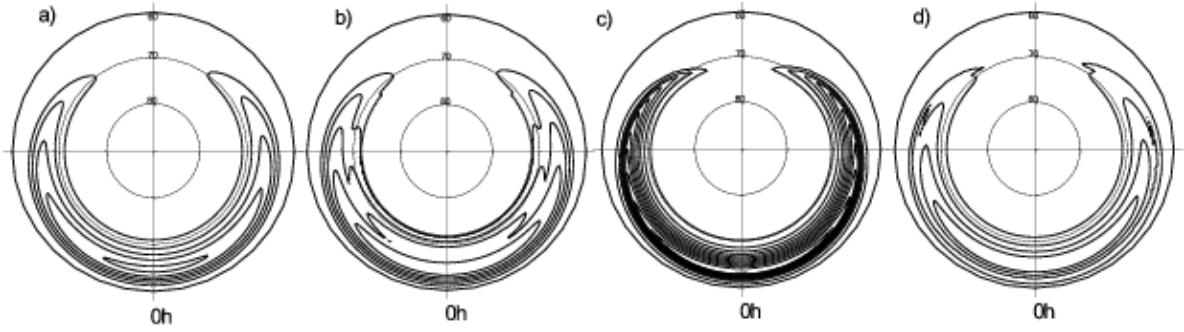

*Fig. 3. Equidensity contours of the precipitating electron flux for unsteady boundary conditions: a) t=0 s, b) t=1000 s, c) t=2800 s, d) t=4500 s.*

Integral conductivity for the Pedersen current is:

$$\Sigma_p = (e^2 n_e/M_i)\int v_{in}/(\omega_{iB}^2 + v_{in}^2)dz, \qquad (3)$$

where **e** is the electron charge, $M_i$ is the ion mass, $\omega_{iB}$ is the ion gyrofrequency, and $v_{in}$ is the ion-neutral collision frequency. The domain of integration is throughout the entire dynamo-region, i.e. from 100 to 120 km.
Observational data and theoretical estimates show that the scale of the electric field along the latitude is several times larger than the scale of precipitation and, hence, than the scale of the conductivity region. Therefore:

$$\partial J/\partial\theta = (\partial\Sigma_p/\partial\theta)E_\theta + (\partial E_\theta/\partial\theta)\Sigma_p \sim E_\theta \, \partial\Sigma_p/\partial\theta \qquad (4)$$

Hence it follows that the electric field configuration is unimportant for the problem of generation of field-aligned currents in the ionosphere, at least for the divergence of those Pedersen currents, which flow along the latitude and produce "curtain" structures. Of importance are the parameters of precipitation, the intensity of which is closely associated with the spatial distribution of the number density of particles in the magnetosphere and, hence, with the pressure relief. For that reason, there must be a correspondence between the picture of field-aligned currents calculated from the gas pressure distribution in the magnetosphere and the picture of field-aligned currents calculated from the distribution of ionization (i.e. precipitation!). It is this reasoning that dictated the formulation of the problem. At the first stage of research, the result of which are presented in this paper, we present the pictures of field-aligned current distribution inferred in terms of a very simple model.
Calculations were performed by the formula:

$$j_{||} \approx j_r = [\partial J_\lambda/\partial\lambda + \cos\theta_a J_\theta + \sin\theta_a \partial J/\partial\theta]/r_o\sin\theta_a \qquad (5)$$

Since in high latitudes the direction of geomagnetic field lines is close to a radial direction, we identified the field-aligned currents in the ionosphere with radial ones. The error arising in this case in the value of the field-aligned current for the auroral zone is less than 20%. Surface densities of Pedersen currents along the latitude, $J_\lambda$, and along the meridian, $J_\theta$, were calculated by standard formulas:

$$J_\lambda = \Sigma_p E_\lambda \quad \text{and} \quad J_\theta = \Sigma_p E_\theta \tag{6}$$

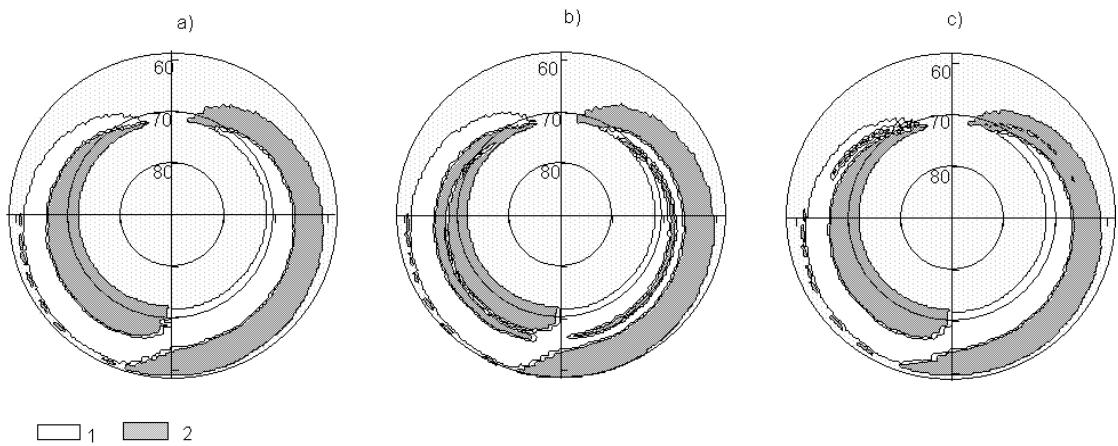

*Fig. 4. Field-aligned currents "generated" in the ionosphere: 1 - zone of inflow currents, 2 - zone of outflow currents; a) t=0 s; b) t=1000 s; c) t=2800 s.*

Results of calculations of field-aligned currents "generated" in the ionosphere are shown in Fig. 4.

It should be noted that only the sign of the field-aligned current whose amplitude exceeded some given value, was plotted. For comparison, for the same instants of time Fig. 5 presents the signs of field-aligned currents generated in the magnetosphere.

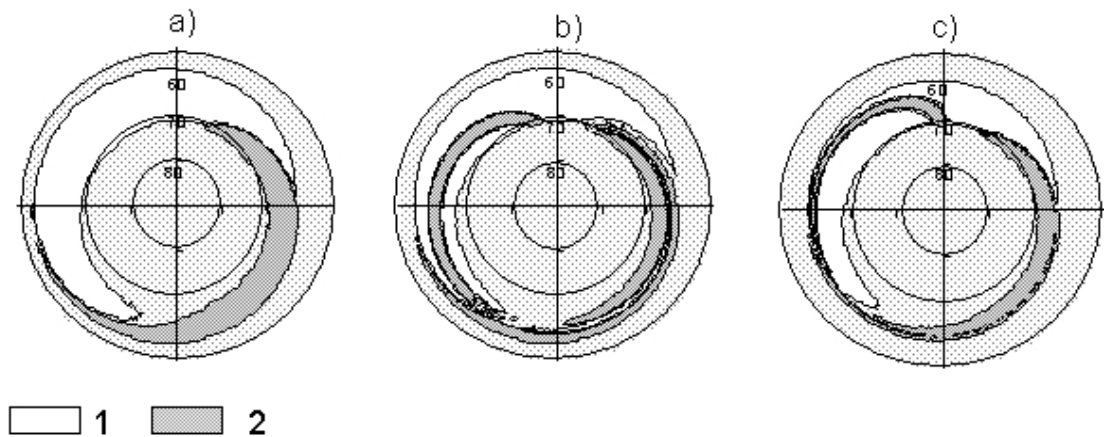

*Fig. 5. Field-aligned currents generated in the magnetosphere: 1 - zone of inflow currents, 2 - zone of outflow currents; a) t=0 s; b) t=1000 s; c) t=2800 s.*

Calculations were performed by the formula:

$$j_{\uparrow\uparrow} = B^i \int \mathbf{j}_\perp \nabla p_B \, dl / p_B B, \quad (7)$$

where the integration domain is along the magnetic field line from the equator to the ionosphere. Magnetic pressure is designated by $p_B$, $\mathbf{dl}$ is the length element of a magnetic field line, and $\mathbf{j}_\perp$ is given by formula (1).

**Discussion of results**

The problem of conjugation of field-aligned currents generated in the magnetosphere, and of field-aligned currents, which are produced as a result of a spatial inhomogeneity of conductivity (and to a lesser extent, of the electric field), that is, as if they were "generated" in the ionosphere, is part of the problem of ionosphere-magnetosphere coupling. It is clear that in actual fact they are simply parts of one and the same global ionospheric-magnetospheric current system. The problem of ionosphere-magnetosphere coupling primarily implies that it is necessary to solve the question as to how the magnetospheric producer of current and power "adjusts itself" to the ionospheric consumer. For a certain special configuration, this problem was solved by P.A. Sedykh [4,5].

It turned out, firstly, that the ionospheric consumer updates the convection rate and through it the plasma pressure gradient, which determines the density of bulk currents which, in turn, determines the behavior of field-aligned currents through its divergence. Secondly, it turned out that ionospheric and magnetospheric currents are not rigidly linked. Some of the current (and power!) that is not "demanded" by the ionosphere can go into feeding the MHD compressor pumping plasma into the region of increase magnetic pressure - in the earthward direction.

For us, the most important issue in this paper is that of ascertaining the direction of the cause-and-effect relationship. Current is primary in the magnetosphere, whereas the electric field is primary in the magnetosphere. Furthermore, the convection system can undergo some adjustment, and together with it the electric field in the ionosphere. But such adjustment is possibly only as corrections of the first approximation to the zero-order approximation. And hence the zero-order approximation, that is, the picture of field-aligned currents obtained essentially for an arbitrary but smooth initial electric field must contain the main elements of the natural system of field-aligned currents which is determine by the distribution of electron precipitation closely associated with the plasma pressure distribution in the magnetosphere. Let us now consider from this standpoint Figs. 4 and 5.

Fig. 4 shows a classical picture of field-aligned currents that coincides in its main traits with the well-known Iijima-Potemra scheme [1]. This correspondence indicates that the factor that determines the main features of the configuration of field-aligned currents is the existence in the ionosphere of a well conducting channel produced by zones of intense precipitation of electrons from the plasma pressure hump region (see Figs. 1 and 3).

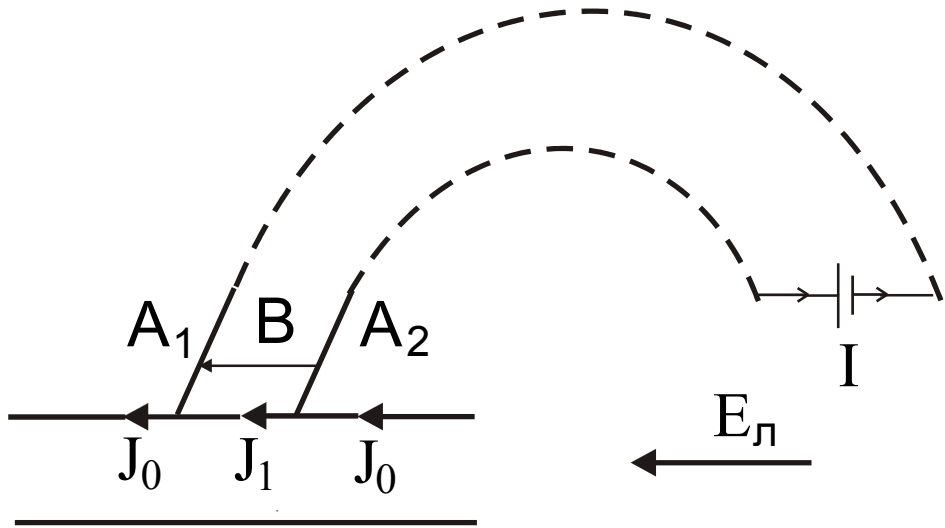

*Fig. 6.* A - ionospheric region with background (low) conductivity. B - ionospheric region with increased conductivity. The portion $A_1$ -$A_2$ has a large-scale (quasi-homogeneous) meridional electric field $E^A_\lambda$ that produces the electric current. If the source of current I in the corresponding magnetospheric region is absent, then the current j flows in the entire portion $A_1$ -$A_2$ ; however, the electric field in region B, $E^B_\lambda$ , is decreased. If, however, the source of current I is present in the magnetosphere, then $E_\lambda$ is everywhere identical, and on portion B the electric current $j^B$ is enhanced. Hence in the former case $j^B = j_0$ and $E^B_\lambda < E^A_\lambda$ , and in the latter case $E^B_\lambda = E^A_\lambda$ , but $j^B = j_0$ .

However, whether or not the enhancement of current in this channel with enhanced conductivity is possible will depend on whether the magnetospheric source is able to supply field-aligned currents this peculiar "discharge gap", as shown in Fig.6.
Fig. 5 shows the picture of field-aligned currents that is "offered" by the magnetosphere. One can see that "demand" and "supply" are more or less identical for the arrangements of the zones. It should be noted that the integral over all inflow and outflow currents in Fig. 5 is virtually zero.

**Conclusion**

The results presented in this study induce us to hope that the conjugation of field-aligned currents of magnetospheric and ionospheric origins is feasible.

*Acknowledgement.* This work was done under project № 02-05-64066, № 03-05-06477.